# Gate-tunable negative differential conductance in hybrid semiconductor-superconductor devices

Mingli Liu,[1, 2, #] Dong Pan,[3, #] Tian Le,[1] Jiangbo He,[1] Zhongmou Jia,[1, 2] Shang Zhu,[1, 2] Guang Yang,[1] Zhaozheng Lyu,[1] Guangtong Liu,[1, 4] Jie Shen,[1, 4] Jianhua Zhao,[3, *] Li Lu,[1, 2, 4, *] and Fanming Qu[1, 2, 4, *]

[1]*Beijing National Laboratory for Condensed Matter Physics, Institute of Physics, Chinese Academy of Sciences, Beijing 100190, China*

[2]*School of Physical Sciences, University of Chinese Academy of Sciences, Beijing 100049, China*

[3]*State Key Laboratory of Superlattices and Microstructures, Institute of Semiconductors, Chinese Academy of Sciences, P.O. Box 912, Beijing 100083, China*

[4]*Songshan Lake Materials Laboratory, Dongguan 523808, China*

[#] These authors contributed equally to this work.

[*] Email: jhzhao@semi.ac.cn; lilu@iphy.ac.cn; fanmingqu@iphy.ac.cn

**Negative differential conductance (NDC) manifests as a significant characteristic of various underlying physics and transport processes in hybrid superconducting devices. In this work, we report the observation of gate-tunable NDC outside the superconducting energy gap on two types of hybrid semiconductor-superconductor devices, i.e., normal metal-superconducting nanowire-normal metal and normal metal-superconducting nanowire-superconductor devices. Specifically, we study the dependence of the NDCs on back-gate voltage and magnetic field. When the back-gate voltage decreases, these NDCs weaken and evolve into positive differential conductance dips; and meanwhile they move away from the superconducting gap towards high bias voltage, and disappear eventually. In addition, with the increase of magnetic field, the NDCs/dips follow the evolution of the superconducting gap, and disappear when the gap closes. We interpret these observations and reach a good agreement by combining the Blonder-Tinkham-Klapwijk (BTK) model and the critical supercurrent effect in the nanowire, which we call the BTK-supercurrent model. Our results provide an in-depth understanding of the tunneling transport in hybrid semiconductor-superconductor devices.**

The research on hybrid semiconductor-superconductor devices based on nanowires with strong spin-orbit coupling (SOC) has been particularly attractive in recent years, due to the underlying exotic physics and potential applications[1-20]. When the system is configured to present topological superconductivity, theories predict the existence of Majorana bound states (MBSs) which could be the building blocks of topological quantum computation[21-24]. Experimentally, measurement of the differential conductance using a tunneling probe or a point contact provides plentiful physical information of the system, such as the superconducting energy gap, Andreev bound states, the strength of the barrier, and the zero-bias conductance peak that may be characteristics of MBSs[25,26]. In addition to the ordinary positive differential conductance, there also exists negative differential conductance (NDC) (or a positive differential conductance dip) in the tunneling spectroscopy or point-contact spectroscopy, reflecting rich physical properties of the device including the superconducting gap, the sub-gap states, the coupling strength, the relaxation of quasiparticles, and the contact conditions, etc.

In the exploration of hybrid superconducting devices, NDC in the tunneling spectroscopy has been investigated in various device geometries, which can be roughly classified to the following three categories. (1) Superconductor-quantum dot-superconductor (S-QD-S) devices [2,9,27-34]. NDC appears in the coulomb diamond with an odd number of electrons, which is caused by the extremely asymmetric coupling strength of the two superconducting electrodes. (2) Majorana island devices[12,35-40]. When the quasiparticle takes a long time to relax to the bound state, the differential conductance $dI/dV$ will be negative (where $I$ is the current, and $V$ is the voltage), i.e., NDC appears. (3) Superconducting tunnel junction with superconductors of different energy gaps ($S_1$-I/N-$S_2$, where I/N is the insulator/normal metal that functions as the tunneling barrier)[41-43]. When the bias voltage increases to be $|\Delta_1 - \Delta_2|/e < V < |\Delta_1 + \Delta_2|/e$ (where $\Delta_1$ and $\Delta_2$ are the superconducting gap of $S_1$ and $S_2$, respectively, and $e$ is the elementary charge), the NDC could be observed due to a sharp decrease in the density of states.

On the other hand, for the measurement of point-contact spectroscopy, the positive differential conductance dip has been extensively reported[44-50] and can be generally attributed to the critical supercurrent effect due to the Maxwell resistance in the non-ballistic regime, as reported by Sheet et al.[45]. Moreover, Shan et al.[44] explained the dip as the combination of superconducting proximity in the normal metal and Josephson tunnelling in the polycrystalline superconductor.

However, NDCs outside the superconducting gap and a gate-tunable evolution from NDC to a differential conductance dip have been elusive. Here, we report the observation of such NDCs and their gate tuning in tunneling spectroscopy of two types of hybrid devices, i.e., the normal metal-superconducting nanowire-normal metal (N-SNW-N) and normal metal-superconducting nanowire-superconductor (N-SNW-S) devices. These NDCs have several distinct characteristics.

First, the amplitude of the NDCs is tunable by gate. The NDC is deep at the high-conductance regime (large gate voltage), and weakens to be a dip (positive value) at the low-conductance regime (small gate voltage). Second, the NDC/dip appears at a bias voltage outside the superconducting gap, and goes away from the gap towards large bias voltage with the decrease of conductance (gate voltage). Third, with the increase of magnetic field, the NDC/dip gradually approaches to zero bias voltage and disappears when the superconducting gap closes. We provide a phenomenological interpretation and achieve a good agreement with the observations, by considering the Blonder-Tinkham-Klapwijk (BTK) model[51] combined with the critical supercurrent in the superconducting nanowire, which is hereinafter referred to as the BTK-supercurrent model.

The superconducting nanowires (SNWs) in our hybrid N-SNW-N and N-SNW-S devices are InAsSb nanowires with an epitaxial superconductor Al layer (~12 nm), grown *in situ* by molecular beam epitaxy method. The ternary compound InAsSb[33,52] is predicted to possess a stronger SOC than its binary compounds InAs and InSb, so that could potentially provide a more promising platform for the research of MBSs. The fabrication details of the superconducting (S) and normal (N) electrodes can be found in the Supplementary material[53]. The junction segment of device A [Fig. 1(a)] is ~10 nm on the left and ~14 nm on the right, and that of device B is ~30 nm (only one junction). The doped silicon substrate with 300 nm-thick $SiO_2$ is used as the global back-gate, and the Ti/Au near the junction segment could be used as the side-gate but are not functional in this work. The differential conductance $dI/dV$ is measured in a dilution refrigerator with a base temperature of ~10 mK by using a standard lock-in-amplifier technique.

We first show the results of device A, which is of N-SNW-N type as shown in Fig. 1(a). Three normal metal electrodes (N) are marked as I, II, and III, respectively. Figure 1(b) shows the differential conductance $dI/dV$ as a function of bias voltage $V$ and back-gate voltage $V_{bg}$ measured by applying a voltage (voltage driven) between electrodes III→I at zero magnetic field. The two horizontal peaks around $V = 0$ represent the superconducting gap, which is harder at larger barrier strength (lower conductance; see Supplementary material[53]). Next, we focus on another obvious feature in Fig. 1(b)—the two blue stripes outside the superconducting gap, which change its color from dark blue to light blue and go to higher bias voltage $|V|$ when decreasing the back-gate voltage from $V_{bg} = 2\ V$ to $V_{bd} = -2\ V$. These stripes disappear at $V_{bg} < -2\ V$ as marked by the yellow arrow. Three typical line-cuts as indicated by the black, red and blue bars are plotted in Fig. 1(c). At $V_{bg} = 2\ V$ (black curve), NDCs outside the superconducting gap can be clearly recognized. At $V_{bg} = 1\,V$ (red curve), the NDCs weaken to be positive differential conductance dips and move towards higher $|V|$. And at $V_{bg} = -1\,V$ (blue curve), the amplitude of the dips decreases and the positions move to higher $|V|$ further.

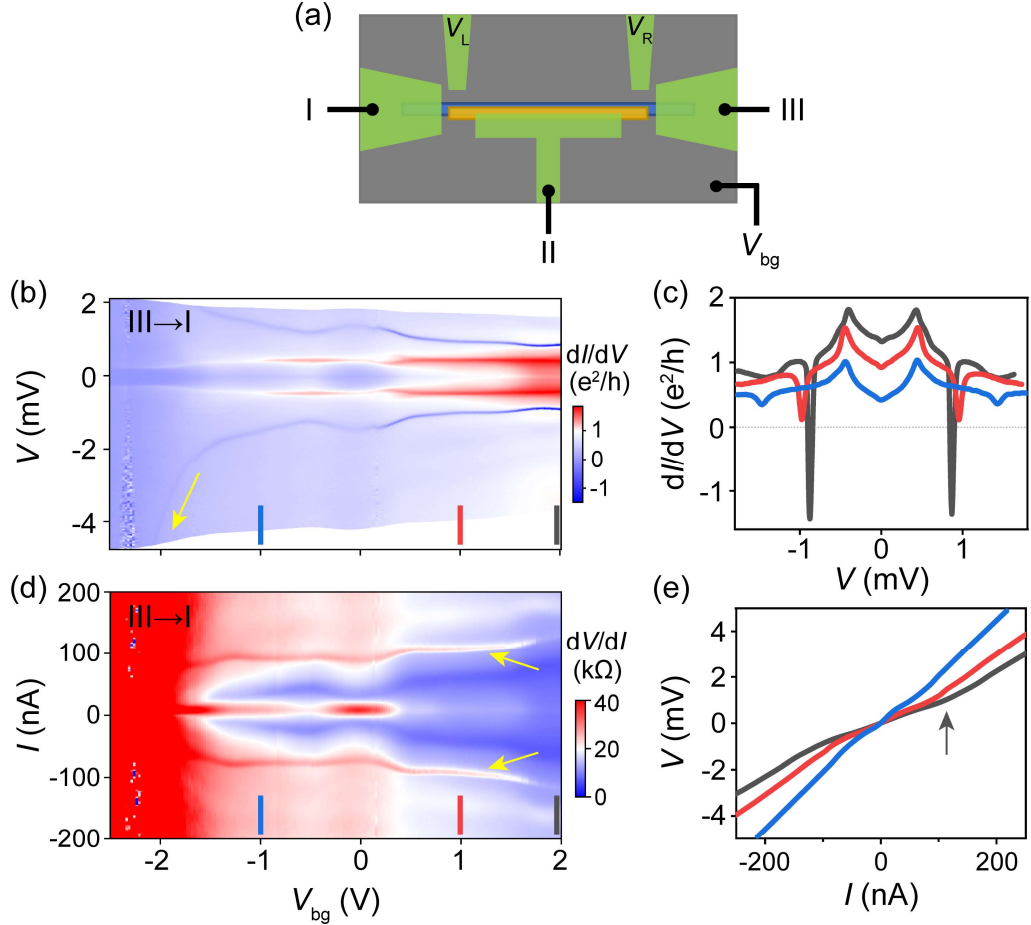

**FIG. 1.** NDCs in device A between electrodes III→I. (a) Schematic diagram of device A with three normal metal electrodes (green) marked as I, II, III, respectively. The nanowire is InAsSb (blue) with epitaxial superconductor Al (yellow). (b) The differential conductance $dI/dV$ as a function of bias voltage $V$ and back-gate voltage $V_{bg}$ for electrodes III→I, i.e., electrode III is the source, and electrode I is the drain. (c) Linecuts taken from (b). (d) The differential resistance $dV/dI$ as a function of $V$ and $V_{bg}$ for electrodes III→I. The two yellow arrows indicate the differential resistance peaks. (e) Integrated linecuts from (d). The black, red and blue curves in (c) and (d) correspond to $V_{bg} = 2, 1, -1\,V$, respectively.

In order to compare with the differential conductance spectroscopy [Fig. 1(b)] and understand the NDCs/dips, we measured the differential resistance $dV/dI$ spectroscopy for the same electrodes III→I in the current-driven mode, as shown in Fig. 1(d). The NDCs/dips in Fig. 1(b) manifest as a pair of differential resistance peaks in Fig. 1(d), as marked by the two yellow arrows, whose occurrence in $I$ could be attributed to the critical supercurrent of the proximity-induced superconducting nanowire (SNW) (as shown later in our BTK-supercurrent model). The three typical line-cuts as indicated by the black, red and blue bars are integrated to obtain the $V - I$ curves, as displayed in Fig. 1(e). The critical supercurrent is roughly estimated to be 120 nA (as indicated by the black arrow) at $V_{bg} = 2\,V$. Note that the NDCs/dips in Fig. 1(b) and the differential resistance peaks in Fig. 1(d) are both tunable by back-gate voltage [the transform from $dI/dV$ to $dV/dI$ of Fig. 1(b) is almost the same as Fig. 1(d); see Supplementary material[53]], and we initially speculate that the occurrence of NDCs/dips is related to the supercurrent (as discussed later).

It is interesting to note that the differential conductance $dI/dV$ spectroscopy for electrodes III→II has no obvious NDCs, but only small positive dips (not reaching negative values), as shown in Figs. 2(a) and (b). However, the evolution of the dips when sweeping $V_{bg}$ is almost the same as that for electrodes III→I [Fig. 1(b)]. In particular, when a perpendicular magnetic field $B_Z$ is applied, the differential conductance dips and correspondingly the differential resistance peaks disappear as the superconducting gap closes, which indicates once again that these features are related to superconductivity, as shown in Figs. 2(c) and (d). (See Supplementary material for the results of electrodes I→II and additional details of device A.)

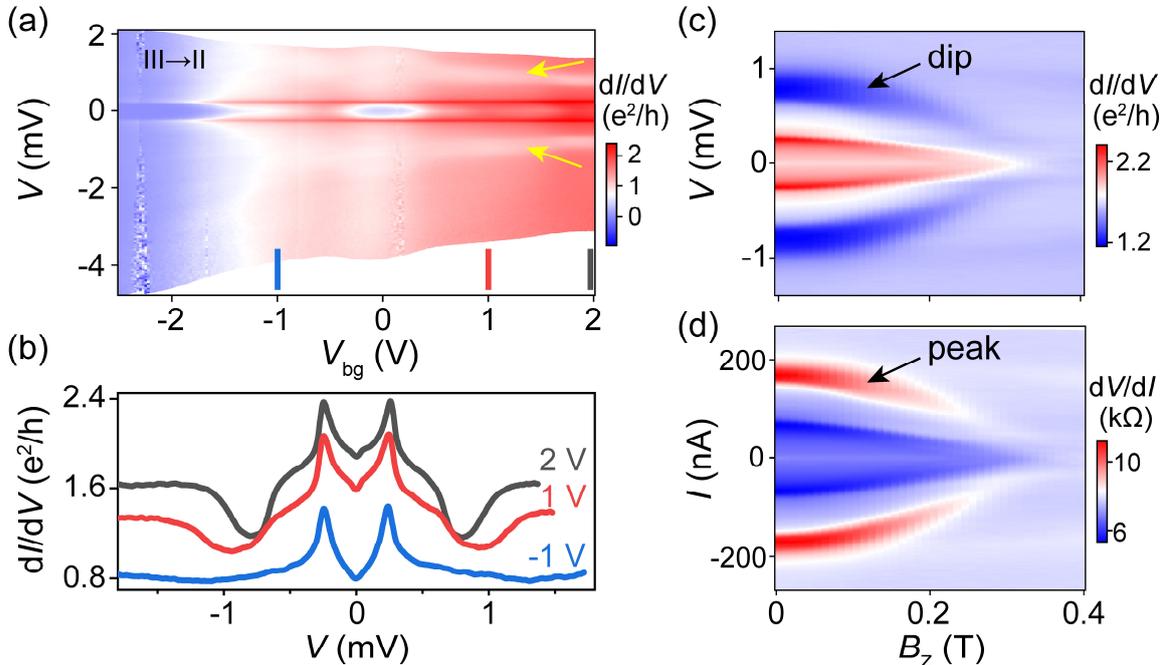

**FIG. 2.** NDCs/dips in device A for electrodes III→II. (a) $dI/dV$ as a function of $V$ and $V_{bg}$ for electrodes III→II. (b) Linecuts taken from (a). (c, d) 2D differential conductance $dI/dV$ and differential resistance $dV/dI$ maps, respectively, showing the evolution of the dips and peaks in magnetic field at $V_{bg} = 2\,V$.

To shed more light on the NDCs/dips and to verify the repeatability, we next present the results on device B, which is of the N-SNW-S type. Figure 3(a) shows a false-colored scanning electron microscope (SEM) image of device B, and Fig. 3(b) is the zoom-in of the red box area in Fig. 3(a), showing that the junction segment is ~30 nm. Figure 3(c) shows the differential conductance at $V_{bg} = 20$ V and 0 V, and NDC appears at $V \approx \pm 0.36$ mV and $\pm 0.6$ mV, respectively. The same as in Figs. 1(b) and 2(a), the NDCs in Fig. 3(e) weaken to be positive dips and go away from the superconducting gap when decreasing $V_{bg}$, as shown by the blue stripes (marked by the two yellow arrows). In the pinch-off regime ($V_{bg} < -8$ $V$), the dips disappear (go towards very large $|V|$), and a hard superconducting gap appears, whose size $\Delta \sim 0.3$ meV. (See Supplementary material for additional details of device B.)

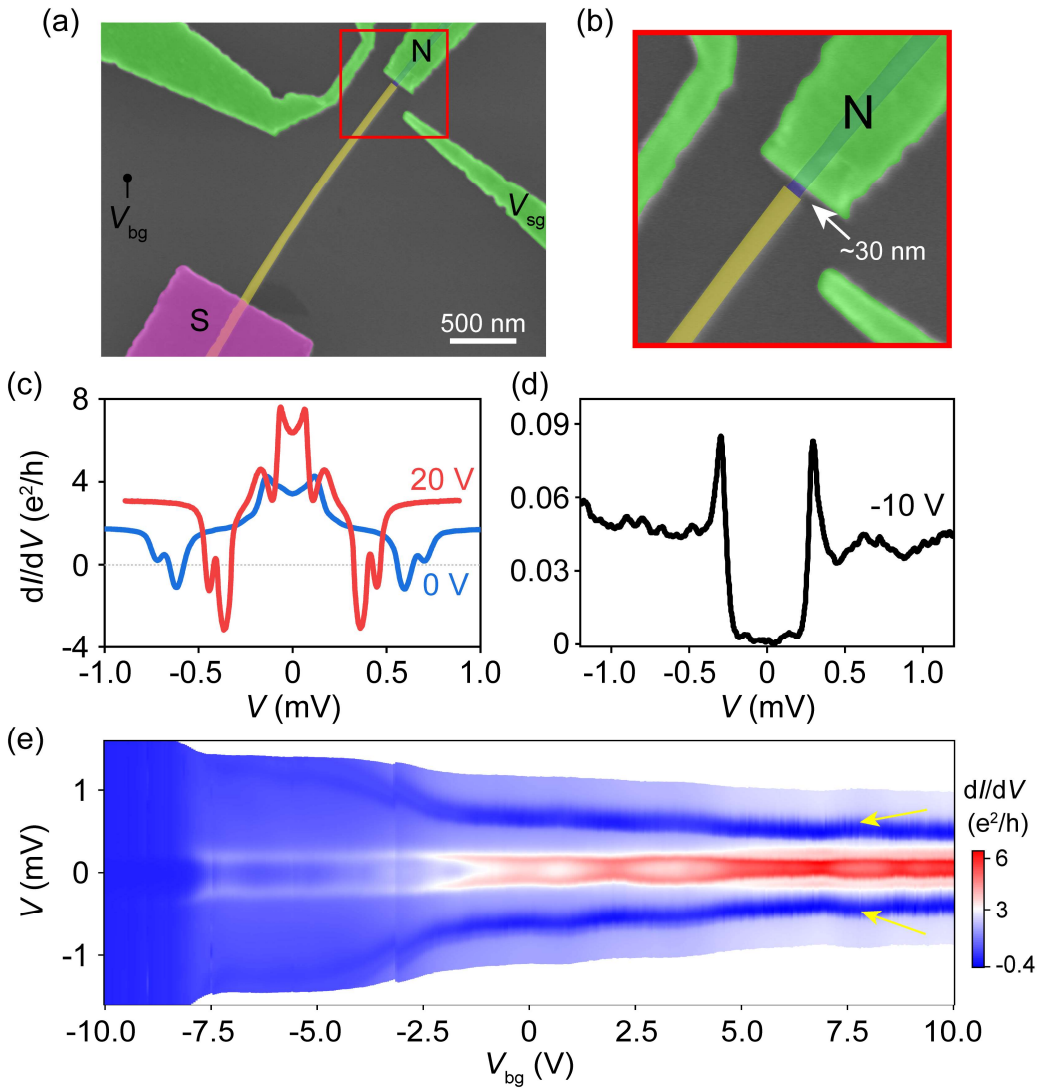

FIG. 3. NDCs in device B. (a) False-colored scanning electron microscope image of device B, where the normal metal (green) electrode is marked as N, and the Al electrode (purple) is marked as S. (b) Red box in (a). The nanowire is InAsSb (blue) with epitaxial superconductor Al (yellow). (c) $dI/dV$ as a function of $V$ at $V_{bg} = 20$ V and 0 V for electrodes N→S. (d) $dI/dV$ shows the superconducting gap at $V_{bg} = -10$ V. (e) 2D $dI/dV$ map as a function of $V$ and $V_{bg}$. Two yellow arrows point to the blue stripes that indicate NDCs/dips.

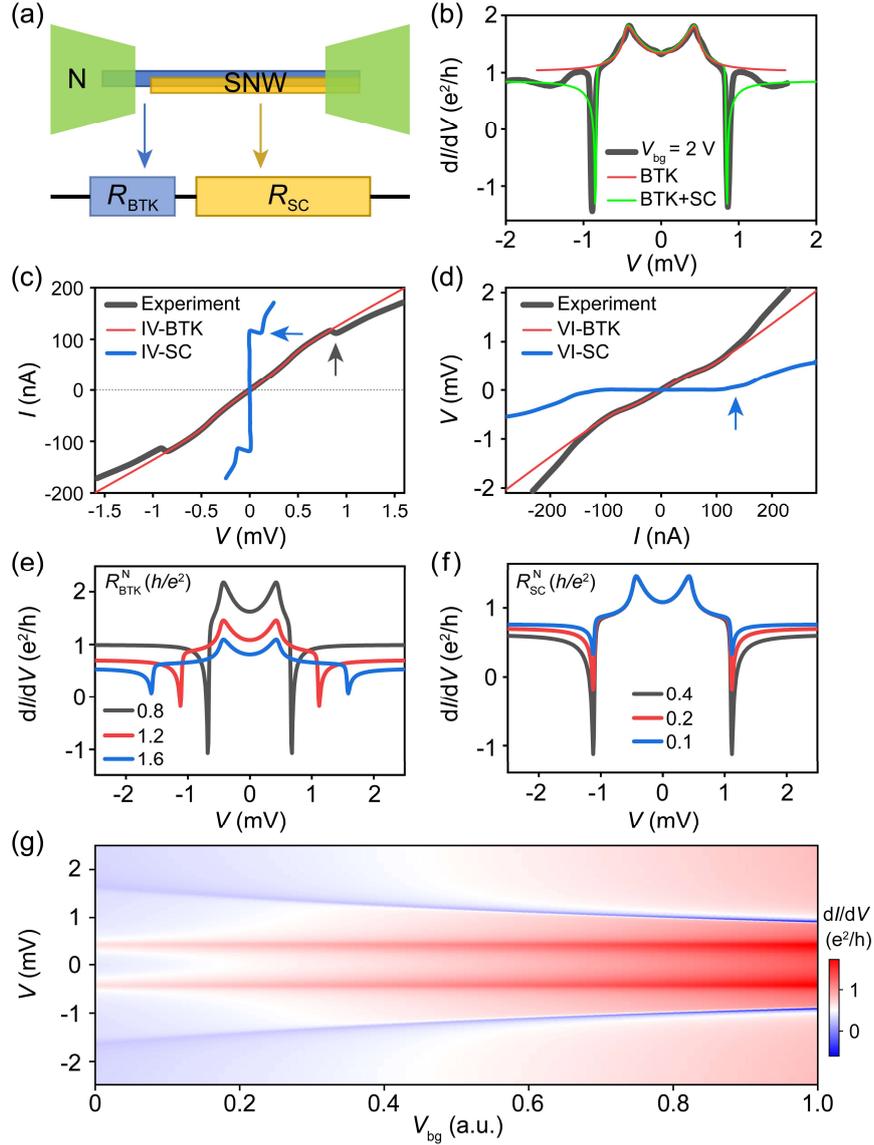

**FIG. 4.** Model and the simulation results of NDCs/dips. (a) Model of a N-SNW junction, which consists of the tunneling resistance $R_{\mathrm{BTK}}$ between N and SNW, and the SNW resistance $R_{\mathrm{SC}}$. (b) $dI/dV$ as a function of $V$ at $V_{\mathrm{bg}} = 2\,V$ for device A (black), the same as the black curve in Fig. 1(c). The red curve is the simulation result using the BTK theory. The green curve is the simulation result using the BTK-supercurrent model. (c, d) $I-V$ and $V-I$ curves (black) of device A for electrodes III→I, obtained by integrating the experimental $dI/dV$ vs. $V$ and $dV/dI$ vs. $I$ curve at $V_{\mathrm{bg}} = 2\,V$, respectively. The red lines are the simulation result using the BTK theory. The blue lines are the result of subtracting the red line by the black line using the current $I$ as the independent variable, respectively. (e) Simulation results for different $R_{\mathrm{BTK}}^N$. Parameters: $\Gamma = 0.03$ meV, $\Delta = 0.46$ meV, $T = 0.03$ K, $Z = 0.4$, $I_c = 120\,nA$, $R_{\mathrm{SC}}^N = 0.2\,h/e^2$, $\gamma_c = 0.01\,nA$. (f) Simulation results for different $R_{SC}^N$. Parameters: $\Gamma = 0.03$ meV, $\Delta = 0.46$ meV, $T = 0.03$ K, $Z = 0.4$, $R_{\mathrm{BTK}}^N = 1.2\,h/e^2$, $I_c = 120\,nA$, $\gamma_c = 0.01\,nA$. (g) Simulation result obtained by assuming that parameters $Z$, $R_{\mathrm{BTK}}^N$, $I_c$, and $\gamma_c$ evolve with $V_{\mathrm{bg}}$ in certain function forms (see Supplementary material).

Gate-tunable NDCs/dips outside the superconducting gap are observed in both types of devices, indicating a universal mechanism. We next combine the BTK model and the critical supercurrent effect (i.e., the BTK-supercurrent model) to understand the behavior of NDCs/dips.

As shown in Fig. 4(a), we model the N-SNW junction in both devices as a series of the tunneling resistance $R_{BTK}$ and the SNW resistance $R_{SC}$. The tunneling resistance $R_{BTK}$ is described by the BTK model, and the current could be calculated as

$$I = 2N(0)ev_F S \int_{-\infty}^{+\infty} [f_0(E - eV) - f_0(E)] [1 + A(E) - B(E)] dE,$$

where $N(0)$ is the density of states at the Fermi level, $S$ is the effective area of the NS interface, $f_0$ is Fermi-Dirac distribution function, $A(E)$ is the probability of Andreev reflection, and $B(E)$ is the probability of normal electron reflection. A one-dimensional potential $V_{NS} = V_0 \delta(x)$ is used to model the NS interface. The barrier strength $Z$ of the NS interface is described by $Z = V_0/\hbar v_F$. We introduce $\Gamma$ to account for the strength of inelastic scattering at the interface, $\Gamma = \hbar/\tau$, where $\tau$ is the lifetime of the quasiparticles. The SNW resistance $R_{SC}$ has typical characteristics of the induced superconducting nanowire, i.e., when the SNW is superconducting, $R_{SC} = 0$; when the current $I$ is greater than the critical supercurrent $I_c$, $R_{SC} = R_{SC}^N$ (the normal-state resistance of the SNW).

The black curve in Fig. 4(b) shows the $dI/dV$ curve of device A for electrodes III→I at $V_{bg} = 2\ V$, which is the same as the black curve in Fig. 1(c). First of all, we use the BTK model to simulate the central segment (the part between the two NDC structures) of the black curve, and plot out the red line using the simulated parameters, as shown in Fig. 4(b). The parameters are the barrier strength $Z = 0.4$, superconducting gap $\Delta = 0.46$ meV, temperature $T = 0.03$ K, the normal-state resistance $R_{BTK}^N = 0.96\ h/e^2$ ($R_{BTK}$ at high-bias voltage), and $\Gamma = 0.03$ meV. Although the simulation could capture the structure inside the energy gap well, the NDCs cannot be simulated. Therefore, it is necessary to consider the critical supercurrent effect and the role of $R_{SC}$ based on BTK theory, i.e., the BTK-supercurrent model (see Supplemental material). The $I - V$ curve incorporating $R_{SC}$ can be expressed as $V_{SC} = R_{SC}^N \sqrt{(I + i\gamma_c)^2 - I_c^2}$, where $\gamma_c$ is used to adjust the broadening near the critical supercurrent resulting from finite temperature and disorder. In order to simulate the NDCs, we consider the voltage-driven measurement (sweeping voltage $V$). The current can be expressed as $I = V/(R_{BTK} + R_{SC})$. $dI/dV$ can be extracted after differentiation, and the contribution of $R_{SC}$ is a key to the NDCs.

Then, we simulate the measured black curve in Fig. 4(b) using the BTK-supercurrent model and obtain the green curve, which shows a good agreement with the experimental result (see Table 2 in the Supplementary material for parameters). To further verify the applicability of our BTK-supercurrent model, we extract the critical supercurrent from both the voltage-driven (sweeping voltage) and current-driven (sweeping current) measurement, as shown in Figs. 4(c) and (d), respectively. The black $I-V$ and $V-I$ curves in Figs. 4(c) and (d) are the integral of the measured d$I$/d$V$ vs. $V$ and d$V$/d$I$ vs. $I$ curve at $V_{bg} = 2\ V$, and the red curves are the integral of the simulated red curve in Fig. 4(b). The $I-V$ and $V-I$ curves of the induced SNW can be obtained by subtracting the red curve from the black curve using $I$ as the independent variable, as shown by the blue curves in Figs. 4(c) and (d). As indicated by the blue arrows, the critical supercurrent is $I_c \sim 120$ nA. As expected, the d$I$/d$V$ spectroscopy and d$V$/d$I$ spectroscopy give the same $I_c$. Remarkably, a sudden current drop as marked by the black arrow in Fig. 4(c) can be clearly recognized. Specifically, in the voltage-driven measurement, when the current is larger than $I_c$, $R_{SC}$ is not zero but jumps to $R_{SC}^N$, so the current decreases, resulting in NDC. In the contrary, for the current-driven measurement, i.e., the d$V$/d$I$ spectroscopy, even if $I$ increases to be larger than $I_c$, the increase of total resistance only enhances the voltage without NDCs, as can be clearly seen from the black $V-I$ curve in Fig. 4(d). What we want to emphasize above is that NDCs could be measured only in the voltage-driven case.

The three experimental characteristics can be well explained by the BTK-supercurrent model. (1) Why do the NDCs weaken with the decrease of $V_{bg}$? This is because the amplitude (depth) of the NDC depends on $R_{BTK}^N$ and $R_{SC}^N$, which are the main parameters that are modified by $V_{bg}$. By fixing other parameters and adjusting $R_{BTK}^N$ or $R_{SC}^N$, respectively, we can simulate the d$I$/d$V$ spectroscopy, as shown in Figs. 4(e) and (f). When $R_{BTK}^N$ decreases or $R_{SC}^N$ increases, the NDCs become deeper. When $V_{bg}$ decreases, $R_{SC}^N$ hardly changes due to the coverage of Al, but $R_{BTK}^N$ increases. Thus, the contribution of $R_{SC}$ to the current reduction is less obvious and the NDC weakens. (2) Why do the NDCs/dips gradually move to the high bias voltage with the decrease of $V_{bg}$? When $R_{BTK}^N$ increases, a higher voltage is required to reach the current that surpasses $I_c$. Figure 4(e) illustrates such behavior clearly. (3) Why do the NDCs/dips follow the evolution of the superconducting gap in a magnetic field? The NDCs/dips originate from the critical supercurrent effect. When the magnetic field increases, the gap and $I_c$ decrease, and eventually NDCs disappear as $I_c$ reaches zero. Regarding the gate tuning, by assuming that parameters $Z$, $R_{BTK}^N$, $I_c$, and $\gamma_c$ evolve with $V_{bg}$ in certain function forms (see Supplementary material), we obtain the $V_\text{bg}$-dependent d$I$/d$V$ spectroscopy, as shown in Fig. 4(g), which is in good agreement with the characteristics of NDCs/dips in our experimental data [Fig. 1(b) and Fig. 3(e)].

One more thing to note is that the NDCs/dips for electrodes III→I are deeper than that for III→II in the same range of $V_\text{bg}$ [see Fig. 1(b) and Fig. 2(a)]. By comparing $R_\text{N}$ (the measured resistance at high bias voltage) at the same $V_\text{bg}$, for example, $R_\text{N} \approx 1.16 \, h/e^2$ (III→I) and $R_\text{N} \approx 0.62 \, h/e^2$ (III→II) at $V_\text{bg} = 2\,V$, and according to the correspondence between a larger $R_\text{SC}^\text{N}$ and a deeper NDC/dip as shown in Fig. 4(f), the reason could be that $R_\text{SC}^\text{N}$ is larger for electrodes III→I due to the longer segment of the nanowire.

In conclusion, we studied NDCs/dips outside the superconducting gap in N-SNW-N and N-SNW-S devices. The observed distinct characteristics as a function of gate voltage and magnetic field could be interpreted by combining the BTK model and the critical supercurrent effect, i.e., the BTK-supercurrent model. Our results provide an in-depth understanding and a reference for the study of the tunnelling spectroscopy of hybrid semiconductor-superconductor devices.

**Acknowledgments**

This work was supported by the National Key Research and Development Program of China (2022YFA1403400 and 2017YFA0304700), by the NSF China (12074417, 92065203, 92065106, 61974138, 11774405, 11527806 and 12104489), by the Strategic Priority Research Program B of Chinese Academy of Sciences (XDB28000000 and XDB33000000), by the Synergetic Extreme Condition User Facility sponsored by the National Development and Reform Commission, and by the Innovation Program for Quantum Science and Technology (2021ZD0302600). D.P. also acknowledges the support from Youth Innovation Promotion Association, Chinese Academy of Sciences (2017156 and Y2021043).

# Supplementary Material for
# Gate-tunable negative differential conductance in hybrid semiconductor-superconductor devices

Mingli Liu,[1,2,#] Dong Pan,[3,#] Tian Le,[1] Jiangbo He,[1] Zhongmou Jia,[1,2] Shang Zhu,[1,2] Guang Yang,[1] Zhaozheng Lyu,[1] Guangtong Liu,[1,4] Jie Shen,[1,4] Jianhua Zhao,[3,*] Li Lu,[1,2,4,*] and Fanming Qu[1,2,4,*]

[1]*Beijing National Laboratory for Condensed Matter Physics, Institute of Physics, Chinese Academy of Sciences, Beijing 100190, China*
[2]*School of Physical Sciences, University of Chinese Academy of Sciences, Beijing 100049, China*
[3]*State Key Laboratory of Superlattices and Microstructures, Institute of Semiconductors, Chinese Academy of Sciences, P.O. Box 912, Beijing 100083, China*
[4]*Songshan Lake Materials Laboratory, Dongguan 523808, China*
[#] These authors contributed equally to this work.
[*] Email: jhzhao@semi.ac.cn; lilu@iphy.ac.cn; fanmingqu@iphy.ac.cn

## 1. Scanning electron microscope (SEM) image of Device A

For both types of devices, the superconducting electrodes (S) are fabricated using standard electron-beam lithography followed by electron-beam evaporation of Al (~80 nm). The normal electrodes (N) are fabricated by selectively etching away the Al layer prior to a direct deposition of Ti/Au (8 nm/80 nm) using a double-layer resist. Short junctions less than 50 nm between N and the superconducting nanowire (SNW) can be realized by utilizing the undercut structure of the double-layer resist and such one-step fabrication process.

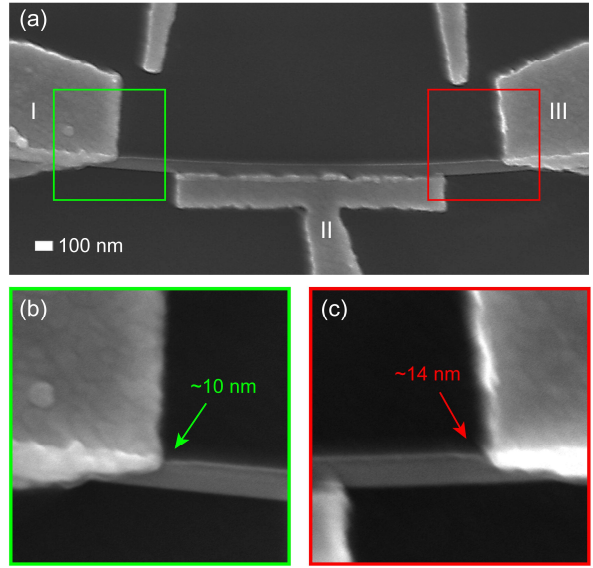

Fig. S1. (a) SEM image of Device A. The corresponding schematic diagram is shown in Fig. 1(a) in the main text. (b, c) Zoom-in of the green and red box area in (a), respectively. The left junction segment is ~10 nm, and the right is ~14 nm.

## 2. Additional data on device A

This section shows additional data on device A.

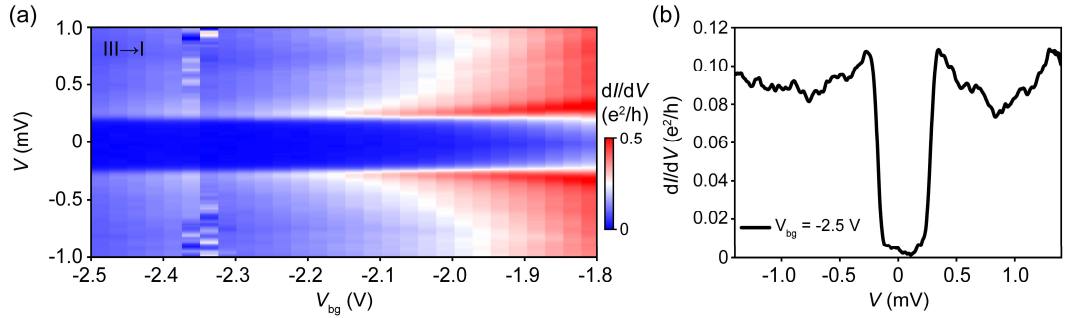

Fig. S2. Additional data on device A for electrodes III → I at larger barrier strength. (a) The differential conductance $dI/dV$ as a function of bias voltage $V$ and back-gate voltage $V_{bg}$ for electrodes III → I. (b) The differential conductance $dI/dV$ linecut at $V_{bg} = -2.5\,\text{V}$. A hard gap[1] can be inferred from the ratio between the normal and superconducting state conductance, $G_N/G_S \sim 90$.

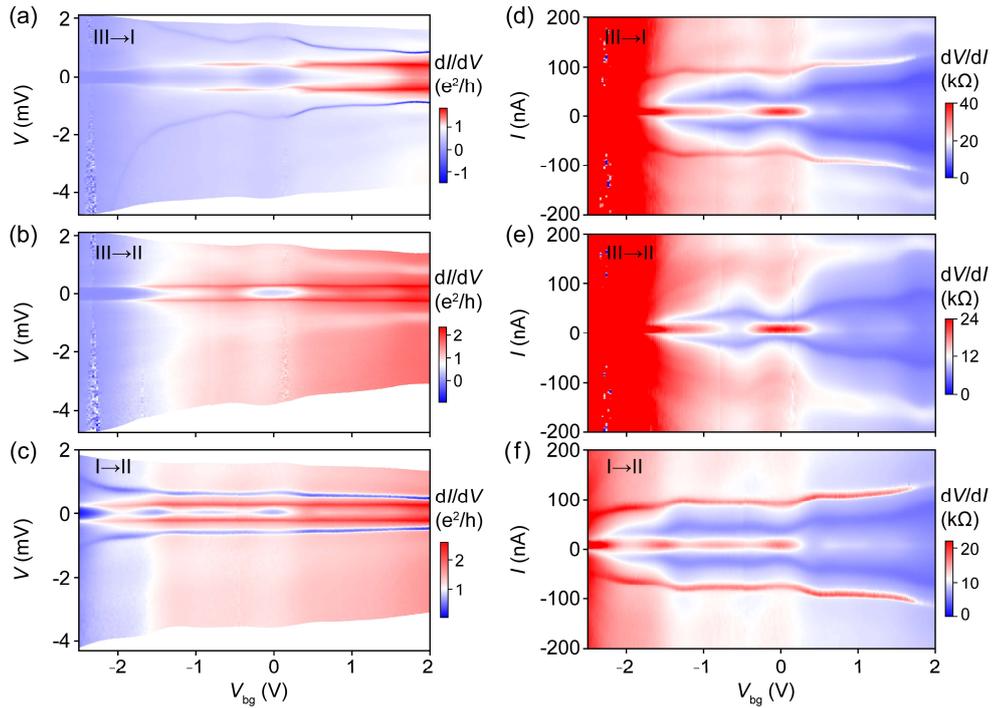

Fig. S3. (a-c) The measured $dI/dV$ spectroscopy in the voltage-driven mode (sweeping voltage) for electrodes III → I, III → II and I → II, respectively. (d-f) The measured $dV/dI$ spectroscopy in the current-driven mode (sweeping current) for electrodes III → I, III → II and I → II, respectively. (a, b, d) The same as Fig. 1(b), Fig. 1(d), and Fig. 2(a), respectively. For clarity and a direct comparison, we plot these three figures here again.

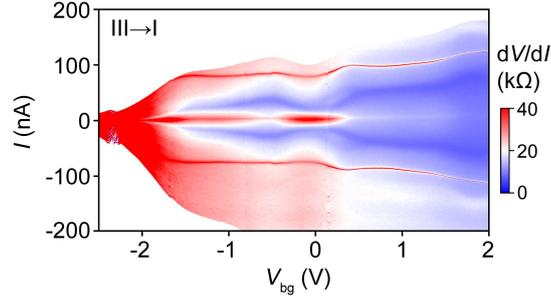

Fig. S4. The result of transforming $dI/dV$ [Fig. 1(b) in main text, i.e., Fig. S3(a)] to $dV/dI$. $I$ is calculated by $\int (dI/dV)dV$. The transformed $dV/dI$ peaks from the voltage-driven measurement show the same behavior as the current-driven measurement, i.e., Fig. S3(d).

### 3. Additional data on device B

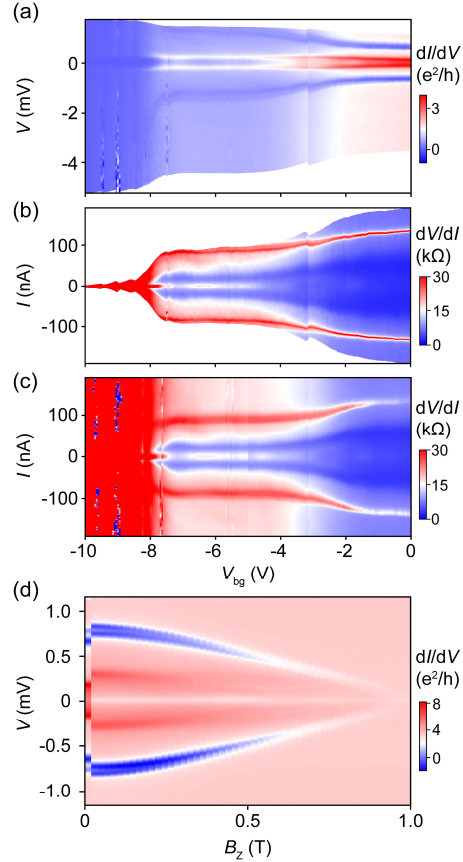

Fig. S5. (a) The measured $dI/dV$ spectroscopy in the voltage-driven mode for device B. The differential conductance dip disappears at $V_{bg} \approx -8$ V. (b) The result of transforming $dI/dV$ in (a) to $dV/dI$. (c) The measured $dV/dI$ spectroscopy in the current-driven mode. (d) 2D $dI/dV$ map showing the evolution of the dip in magnetic field at $V_{bg} = -2$ V. The conductance jump near zero magnetic field is caused by the quench of evaporated Al.

## 4. Details of the theoretical simulation

As explained in the main text, the transport of our semiconductor-superconductor hybrid devices is described by the BTK-supercurrent model. According to the BTK theory[2], the normal metal-superconductor (NS) interface potential barrier is assumed to be a one-dimensional delta function $V_{NS} = V_0 \delta(x)$. When a voltage is applied, the current could be calculated as

$$I = 2N(0)ev_F S \int_{-\infty}^{+\infty} [f_0(E-eV) - f_0(E)][1 + A(E) - B(E)]dE,$$

where $N(0)$ is the density of states at the Fermi level, $S$ is the effective area of the NS interface, $f_0$ is the Fermi-Dirac distribution function, $f_0(E-eV) = \left[1 + \exp(\frac{E-ev}{k_B T})\right]^{-1}$, $k_B$ is Boltzmann constant, $T$ is temperature. $A(E)$ is the probability of Andreev reflection, $A = u_0^2 v_0^2 / \gamma^2$, and $B(E)$ is the probability of normal electron reflection, $B = (u_0^2 - v_0^2)^2 Z^2 (1+Z^2)/\gamma^2$, where $u_0^2 = 1 - v_0^2 = \frac{1}{2}\{1 + [(E^2 - \Delta^2)/E^2]^{1/2}\}$, $\gamma^2 = \left[u_0^2 + Z^2(u_0^2 - v_0^2)\right]^2$, and $Z = V_0/\hbar v_F$ is a dimensionless parameter that represents the barrier strength. (When $Z = 0$, the barrier is fully transparent and $A = 1$.) The differential conductance of the NS interface can be expressed as

$$\frac{dI}{dV} = 2N(0)ev_F S \int_{-\infty}^{+\infty} \frac{\partial f_0(E-eV)}{\partial f_0(E)}[1 + A(E) - B(E)]dE.$$

Considering the inelastic scattering of the interface, the Bogoliubov coherence factors $u_0$ and $v_0$ need to be rewritten as

$$u_0^2 = \frac{1}{2}\left[1 + \frac{\sqrt{(E+i\Gamma) - \Delta^2}}{E+i\Gamma}\right] = 1 - v_0^2,$$

where $\Gamma$ is the strength of the inelastic scattering, $\Gamma = \hbar/\tau$, and $\tau$ is the lifetime of the quasiparticles. When $\Gamma = 0$, no inelastic scattering occurs at the NS interface. However, with the increase of $\Gamma$, the conductance peaks at the superconducting gap edges as described by the BTK model broaden. The experimental data within the superconducting energy gap can be simulated well by using parameters $\Gamma$, $\Delta$, $T$, $Z$, and $R_{BTK}^N$. The resistance $R_{BTK}^N$ is the $R_{BTK}$ at high bias voltage, used to match the real resistance in the data.

However, the NDCs and the differential conductance dips cannot be simulated by BTK model. Therefore, we add the external supercurrent part on the basis of BTK model, mainly considering the critical supercurrent effect of SNW (superconducting nanowire), and we call it the BTK-supercurrent model. Assuming that $R_{SC}$ is the resistance of the SNW, when the SNW is superconducting, $R_{SC}=0$; when the current $I$ is greater than the critical supercurrent $I_c$, $R_{SC}=R_{SC}^N$ (the normal-state resistance of the SNW). The $I-V$ function of the SNW part can be written as $V_{SC}=R_{SC}^N\sqrt{I^2-I_c^2}$ [3]. Considering the finite temperature and disorder, $I+i\gamma_c$ is used to replace $I$ to adjust the broadening near the critical supercurrent, and thus $V_{SC}=R_{SC}^N\sqrt{(I+i\gamma_c)^2-I_c^2}$. The total resistance can be written as: $R_{tol}=R_{BTK}+R_{SC}$. Taking the parameters of the superconductor part, i.e., $I_c$, $R_{SC}^N$, $\gamma_c$, into account, and combining with the BTK parameters, $\Gamma$, $\Delta$, $T$, $Z$, $R_{BTK}^N$, we can simulate our experimental results.

For Fig. 4(g) in the main text, some parameters are rewritten as a function of $V_{bg}$ to simulate the variation trend with $V_{bg}$. Since we cannot determine the exact relations between these parameters and $V_{bg}$, we just assume function forms phenomenologically, as shown in Table 1.

| Part: $R_{BTK}$ | | Part: $R_{SC}$ | |
|---|---|---|---|
| $\Gamma$ (meV) | 0.03 | $I_c$ (nA) | $0.3*\exp(V_{bg})+0.4$ |
| $\Delta$ (meV) | 0.46 | | |
| $T$ (K) | 0.03 | $R_{SC}^N$ ($h/e^2$) | 0.2 |
| $Z$ | $2/(V_{bg}+4)$ | | |
| $R_{BTK}^N$ ($h/e^2$) | $1/\exp(V_{bg}-1)$ | $\gamma_c$ (nA) | $0.008*\exp(1-V_{bg})$ |

Table 1: Parameters used for Fig. 4(g) in the main text.

Using our BTK-supercurrent model, the three curves in Fig. 1(c) in the main text can be simulated well, as shown in Fig. S6 (see Table 2 for corresponding parameters).

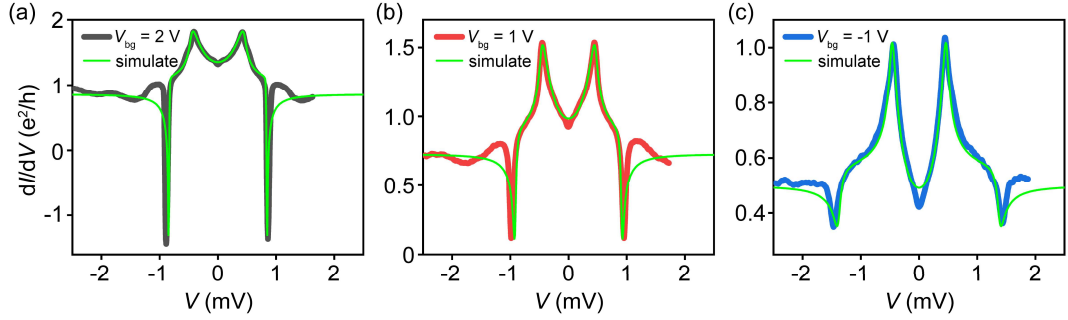

Fig. S6. Simulation of the experiment data shown in Fig. 1(c) in the main text. The measured $dI/dV$ vs. $V$ curves at $V_{bg} = 2\,V$, $1\,V$, $-1\,V$ correspond to the black, red, blue linecuts, respectively. The green lines are the simulation results using our BTK-supercurrent model.

| Parameters | Part: $R_{BTK}$ | | | | | Part: $R_{SC}$ | | |
| --- | --- | --- | --- | --- | --- | --- | --- | --- |
| | $\Gamma$ (meV) | $\Delta$ (meV) | $T$ (K) | $Z$ | $R_{BTK}^N$ ($h/e^2$) | $I_c$ (nA) | $R_{SC}^N$ ($h/e^2$) | $\gamma_c$ (nA) |
| $V_{bg} = 2\,V$ | 0.03 | 0.46 | 0.03 | 0.4 | 0.96 | 120 | 0.2 | 0.002 |
| $V_{bg} = 1\,V$ | 0.03 | 0.47 | 0.03 | 0.48 | 1.16 | 103 | 0.2 | 0.018 |
| $V_{bg} = -1\,V$ | 0.03 | 0.475 | 0.03 | 0.64 | 1.78 | 90 | 0.2 | 0.03 |

Table 2: Parameters used for Figs. S6(a-c).

Supplemental references